\documentclass[review,preprint]{elsarticle}

\usepackage{amsmath}
\usepackage{amsfonts}
\usepackage{array}

\usepackage{lineno,hyperref}

\journal{}

\begin{document}

\begin{frontmatter}

\title{Data-driven Chaos Indicator for Nonlinear Dynamics and Applications
  on Storage Ring Lattice Design}

\author[bnl]{Yongjun Li\corref{correspondingauthor}}
\cortext[correspondingauthor]{Corresponding author}
\ead{yli@bnl.gov}

\author[ihep,ucas]{Jinyu Wan}
\author[purdue]{Allen Liu}
\author[ihep,ucas]{Yi Jiao}
\author[bnl]{Robert Rainer}

\address[bnl]{Brookhaven National Laboratory, Upton, New York 11973, USA}
\address[ihep]{Institute of High Energy Physics, Beijing 100049, China}
\address[ucas]{University of Chinese Academy of Sciences, Beijing 100049,
  China}
\address[purdue]{Department of Electrical and Computer Engineering, Purdue
  University, West Lafayette, Indiana 47907, USA}

\begin{abstract}
  A data-driven chaos indicator concept is introduced to characterize the
  degree of chaos for nonlinear dynamical systems. The indicator is
  represented by the prediction accuracy of surrogate models established
  purely from data. It provides a metric for the predictability of
  nonlinear motions in a given system. When using the indicator to
  implement a tune-scan for a quadratic H\'enon map, the main resonances
  and their asymmetric stop-band widths can be identified. When applied to
  particle transportation in a storage ring, as particle motion becomes
  more chaotic, its surrogate model prediction accuracy decreases
  correspondingly. Therefore, the prediction accuracy, acting as a chaos
  indicator, can be used directly as the objective for nonlinear beam
  dynamics optimization. This method provides a different perspective on
  nonlinear beam dynamics and an efficient method for nonlinear lattice
  optimization. Applications in dynamic aperture optimization are
  demonstrated as real world examples.
\end{abstract}

\begin{keyword}
data-driven chaos indicator\sep surrogate model\sep nonlinear dynamics\sep
dynamic aperture
\end{keyword}

\end{frontmatter}

\nolinenumbers

\section{\label{sect:intro}Introduction}

  It is well-known that the predictability of motion in a nonlinear
  dynamical system is closely associated with its degree of chaos. Given
  an initial condition, although its motion is deterministic, its
  long-term prediction might not be quantitatively accurate because
  numerical errors can be cumulative and amplified. The Lyapunov
  exponent~\cite{vulpiani2009}, i.e., the exponential growth of separation
  of infinitesimally close trajectories, is often used as a chaos
  indicator to characterize the sensitivity of chaotic motion to its
  initial condition.
  
  Consider a different scenario: an unknown nonlinear dynamical system is
  encapsulated into a blackbox and only an ensemble of trajectories (input
  and output data) are available. Comparing actual trajectories to
  interpolated trajectories is one way to gauge chaos. A typical method to
  interpolate from known trajectories is to build a surrogate model with
  machine learning techniques.  A surrogate model needs to be established
  first, then predictions can be made by evaluating trajectories with
  given initial conditions. This procedure is known as ``supervised
  learning''~\cite{mitchell1997}. To validate the model, the data is often
  randomly split into two clusters: a large training set and a small
  testing set. A model is then constructed from the training set. The
  performance of the model, i.e., the prediction accuracy, is measured by
  comparing the testing data against its prediction. The performance of
  the model depends on the type and complexity of the model, the volume of
  training data, the algorithm used for training, etc.  Nevertheless, the
  prediction accuracy depends greatly on the degree of chaos. Therefore,
  an intuitive method for detecting chaos directly, purely from data is
  possible. In other words, predictability itself can act as a chaos
  indicator. From our studies we observed that by using the predictability
  of less-complex surrogate models, and a small volume of training data,
  some nonlinear behaviors in a dynamical system can be well
  characterized.

  Surrogate models have been widely used in studying nonlinear dynamical
  systems~\cite{barahona1996,Schreiber1996,tokuda2001,deshmukh2013,
    lindhorst2014,han2021}, including charged particle motion in modern
  accelerators~\cite{sanchez2017,wang2019,huang2019,Edelen2020,Wan2020,
    Kran2021,zhu2021}. These models are obtained by training on either
  simulated data or experimental data, which have a high computational
  demand or require complicated experimental processing. If models can
  predict the dynamical system properties accurately with reduced resource
  requirements, they can be used for more efficient applications, such as
  optimization problems. Improving the prediction accuracy is the highest
  priority in these applications. In contrast to these existing
  approaches, the main advantage of using data-driven chaos indicators is
  that the requirement on the absolute accuracy of surrogate models is
  less demanding, and therefore can be structured with less complexity and
  data.

  To further explain this approach, the remaining sections are outlined as
  follows: in Sect.~\ref{sect:henon}, the detailed data-driven chaos
  indicator concept is introduced and applied to implement a tune-scan for
  a quadratic H\'enon map. Sect.~\ref{sect:performanceChaos} correlates
  the nonlinear dynamical behavior of charged particles circulating in a
  storage ring with the data-driven chaos indicator, constructed with an
  artificial neural network (ANN)~\cite{hassoun1995}. In
  Sect.~\ref{sect:application}, we demonstrate the application of chaos
  indicators to optimize storage ring dynamic apertures.  The model
  selection problem and other related issues are discussed in
  Sect.~\ref{sect:selection}. A brief summary is given in
  Sect.~\ref{sect:summary}.

\section{\label{sect:henon}Data-driven chaos indicator for H\'{e}non map}
  
\begin{figure}[!ht]
    \centering
    \includegraphics[width=0.5\columnwidth]{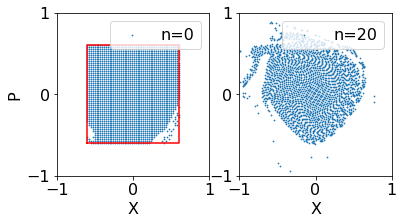}
    \caption{\label{fig:henon} A short-term (20 turns) coordinate
      transformation of the H\'enon map with $\nu=0.205$. The initial
      conditions (left) are populated within the red square
      $(x,p)\in[-0.6,+0.6]$. The blank points represent unstable
      trajectories, where at least one of their $x$ and $p$ amplitudes
      exceed a threshold $10$ during the transformation. The transformed
      coordinates that survived after 20 turns are shown in the right
      subplot.}
  \end{figure}

  The well-studied quadratic H\'enon map, as shown in
  Eq.~\eqref{eq:henon}, is used as an example to demonstrate how to
  construct a data-driven chaos indicator for tune-scanning.  It
  represents a thin sextupole kick followed by a linear rotation in a
  2-dimensional phase space,
  \begin{equation}\label{eq:henon}
    \left(\begin{array}{c}
      x\\
      p
    \end{array}\right)_{n+1}=\left(\begin{array}{cc}
      \cos2\pi\nu & \sin2\pi\nu\\
     -\sin2\pi\nu & \cos2\pi\nu
    \end{array}\right)\left(\begin{array}{c}
      x\\
      p-\lambda x^2
    \end{array}\right)_n,
  \end{equation}
  where, $n$ is a non-negative integer, $\nu$ is known as the linear tune
  of the transformation, and the sextupole strength $\lambda$ is set as
  one for this demonstration. We assume the map in encapsulated as a
  blackbox with its tune as the control knob. For a given tune, as
  illustrated in Fig.~\ref{fig:henon}, some known Trajectories that start
  with initial conditions $(x_0,p_0)$ (input data) within a specific area
  $x_0,p_0\in[-0.6,+0.6]$, end with $(x_n,p_n)$ (output data) after a
  limited number of turns (such as $n=20$). Based on the data, we can
  extract some parameters to characterize its long-term stability such as,
  the location of resonance lines and their stop-band widths, the relative
  size of the stable region, etc. This is accomplished by carrying out a
  tune-scan.

\begin{figure}[!ht]
    \centering
    \includegraphics[width=0.5\columnwidth]{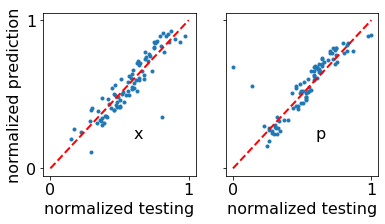}
    \caption{\label{fig:henonDdci} Prediction performance of
      $7^{th}$-order polynomials surrogate model for the H\'enon map in
      the $x$ and $p$ plane at a given tune $\nu=0.205$. Each dot
      represents one initial condition in the testing cluster. The red
      dashed line is the desired/expected value. Note the original data
      has been scaled/normalized to a range of $[0,1]$.}
  \end{figure}

  A tune-scan can be used to compare a nonlinear system's behavior at
  different linear tunes.  At each given tune, some trajectories are
  produced from the blackbox. Most of them (around 80--90\%) are used to
  train a surrogate model with a polynomial regression algorithm,
  \begin{equation}\label{eq:ploy}
    \begin{pmatrix}
      x \\
      p
    \end{pmatrix}_n
    =
    \begin{pmatrix}
      a_0 & a_1 & a_2 & a_3 & a_4 \cdots\\
      b_0 & b_1 & b_2 & b_3 & b_4 \cdots
    \end{pmatrix}
    \begin{pmatrix}
      1   \\
      x   \\
      p   \\
      x^2 \\
      xp  \\
      p^2 \\
      \vdots
    \end{pmatrix}_0.
  \end{equation}
  Here, the $7^{th}$ order polynomials are used. The rest, 10--20\% of the
  data, is used as a testing set for performance validation. The
  validation is done by comparing the testing data against their model
  predictions (Fig.~\ref{fig:henonDdci}). Quantitatively, the prediction
  accuracy is measured with the mean squared errors (MSE) between the
  predictions and the true values. It also serves as the data-driven chaos
  indicator.
  
\begin{figure}[!ht]
    \centering
    \includegraphics[width=0.5\columnwidth]{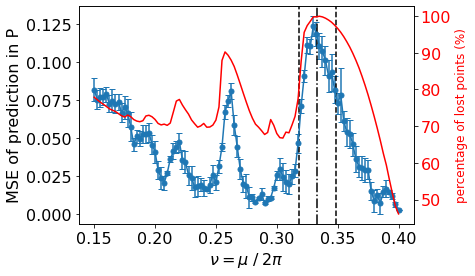}
    \caption{\label{fig:henonTuneScan} Prediction performance of a
      polynomial surrogate model (blue line with error bars) vs. loss rate
      (solid red) of H\'enon map at different tunes. The black dot-dash
      line denotes the location of the $\frac{1}{3}$ resonance line and
      two dashed lines mark out its asymmetric stopband widths at each
      side with a tune separation of $\pm0.015$.}
  \end{figure}

  The tune-scan result is illustrated in Fig.~\ref{fig:henonTuneScan}, in
  which the model's prediction performance is shown as the blue line with
  error bars. For each tune, the data was re-sampled randomly into
  different training/testing sets, multiple times (also known as the
  cross-validation technique). The shuffling of data can avoid selecting
  data that is trapped in a specific resonance, preventing the degree of
  chaos from being under- or over-estimated. The error bars represent the
  statistical fluctuations with different re-samplings.  Due to quadratic
  perturbation, the worst model prediction occurs at $\nu=\frac{1}{3}$ as
  expected, which corresponds to a strong $3^{rd}$-order resonance
  line. This resonance also has the widest stop-band width (approximated
  by the width of half-height of peak). Besides $\frac{1}{3}$, some other
  high order resonances at $\nu=\frac{1}{4}, \frac{1}{5}$, even
  $\frac{1}{7}$ are visible with this chaos indicator. For comparison, a
  long-term (2,048 turns) transformation starting from a wide initial
  condition of the $x$ and $p$ within $[-1.2, +1.2]$ was computed. Its
  loss (i.e., unstable trajectory) rate as the function of the tune is
  also shown as the red solid line in Fig.~\ref{fig:henonTuneScan}. The
  data-driven chaos indicator observed appears to be highly correlated
  with the loss rate of the long-term tracking.

\begin{figure}[!ht]
    \centering
    \includegraphics[width=0.5\columnwidth]{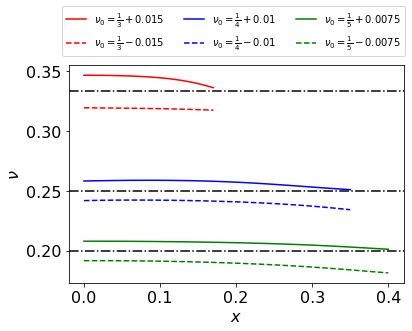}
    \caption{\label{fig:henonStopBand} Different behaviors of the
      amplitude-dependent detuning when the linear tune approaches the
      $\frac{1}{3},\frac{1}{4},\frac{1}{5}$ resonances, which results in
      asymmetric stop-band widths. The solid lines indicate when the
      linear tune is slightly above a resonance, and the
      amplitude-dependent tunes merge to the resonance line (the dot-dash
      lines) quickly. The dashed lines are for the linear tune below the
      resonances, in which amplitude-dependent tune drifts away from the
      resonance. This effect leads to asymmetric and shifted peaks around
      the resonances. The asymmetric detuning effect can be detected by
      both the data-driven chaos indicator and a detailed loss rate
      tracking in Fig.~\ref{fig:henonTuneScan}.}
  \end{figure}

  It is interesting to note that an asymmetric stop-band width is
  detectable with this chaos indicator in
  Fig.~\ref{fig:henonTuneScan}. The appearance of the asymmetry is due to
  amplitude dependent detuning.  It behaves differently when the linear
  tune is slightly off the resonance line as shown in
  Fig.~\ref{fig:henonStopBand}. When the linear tune is below the
  $(\frac{1}{3})^-$ resonance (in the left side), its amplitude-dependent
  tunes drift away from the resonance (dashed line). Therefore, the
  motions are less chaotic, and the left stopband width is narrow. But
  when the linear tune is above the $(\frac{1}{3})^+$ (in the right side),
  its amplitude-dependent tune merges to the resonance quickly (solid
  line), so the motions are more chaotic, and the stopband width at the
  right side is correspondingly wide. This asymmetry is also observed at
  $\nu=\frac{1}{4},\frac{1}{5}$ and confirmed with the loss rate.
  
  The two tune-scan results in Fig.~\ref{fig:henonTuneScan} are closely
  correlated. The information needed for these, however, can be very
  different. Using the data-driven chaos indicator, even short-term (20
  turns) map transformations for only partial initial conditions can
  provide some useful information. The tune-scan using the loss rate is
  more accurate, but it also requires a greater number (2,048 turns) of
  map transformations for more initial conditions of $x$ and $p$. In
  real-world applications, there may be a high resource demand to obtain
  such data. Using limited data resources to obtain an early chaos
  indicator has the potential advantage of boosting the optimization of
  design of a nonlinear dynamical system.

\section{\label{sect:performanceChaos}Data-driven chaos indicator for storage rings}
  
  When studying the stability of charged particles in modern storage
  rings, implementation of particle tracking is often necessary. To
  determine the dynamic aperture (DA) boundary, in which particle motion
  is stable would require this type of simulation. During the design
  stage, many time-consuming tracking simulations are needed to search for
  magnet lattices with sufficient DA. To boost the search process,
  surrogate models have been used to replace the original accelerator
  lattices~\cite{wang2019,Edelen2020,Wan2020,Kran2021}. The accuracy of
  the models usually plays a critical role in ensuring optimization
  converges in the desired direction. Modern accelerators are integrated
  with some strong nonlinear magnets. This can make reliable surrogate
  models difficult to construct. From another perspective, as mentioned
  previously, the prediction accuracy of surrogate models can be a metric
  for the chaos of particle motion.

\begin{figure}[!ht]
    \centering
    \includegraphics[width=0.5\columnwidth]{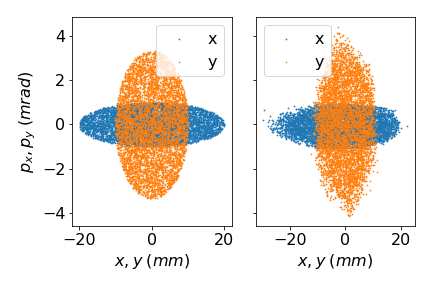}
    \caption{\label{fig:inout} Simulated one-turn transportation of a
      storage ring lattice with some given inputs (left) and their outputs
      (right) in the 4-dimensional phase space. The ranges of inputs are
      comparable to the desired dynamic aperture there. This is the input
      and output data that the ANN is trained on.}
  \end{figure}

\begin{figure}[!ht]
    \centering
    \includegraphics[width=0.5\columnwidth]{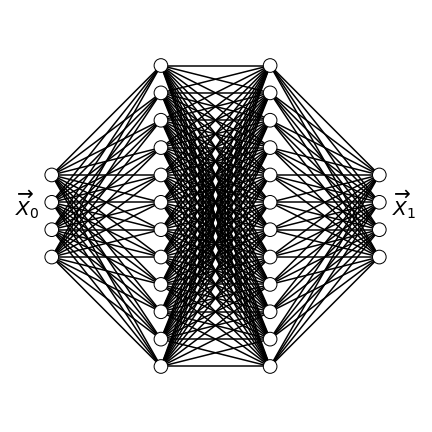}
    \caption{\label{fig:nn} A four-layer $4\times12\times12\times4$ ANN
      used as the surrogate model to represent a storage ring's one-turn
      transportation in the 4-dimensional phase space.}
  \end{figure}

  Consider a storage ring accelerator composed of various magnetic
  elements, in which the transportation of a charged particle for single
  turn (or a few repetitive turns) can be represented by a nonlinear
  transformation
  \begin{equation}
    \overrightarrow{X}_1 = M_{0\rightarrow1} \cdot \overrightarrow{X}_0.
  \end{equation}
  Here, $\overrightarrow{X}_1$, $\overrightarrow{X}_0$ are the particle
  coordinates in the phase space, and $M_{0\rightarrow1}$ is the one-turn
  transportation map. Given the ring magnetic lattice, and using some
  simulated trajectory data $(\overrightarrow{X}_0,\overrightarrow{X}_1)$
  as illustrated in Fig.~\ref{fig:inout}, a surrogate model constructed
  with an artificial neural network (ANN) as shown in Fig.~\ref{fig:nn}
  has been adopted. The reason for using a different model here is to
  indicate that a data-driven chaos indicator represents the
  predictability of the system regardless of the selection of models. For
  this investigation, a simulated trajectory dataset was divided into a
  training set and a testing set. The accuracy of the model was still
  measured by the mean squared error (MSE) between the testing set and its
  model prediction.

\begin{figure}[!ht]
    \centering
    \includegraphics[width=0.49\columnwidth]{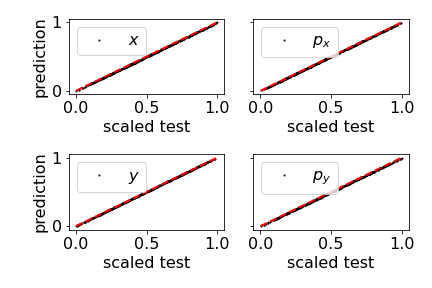}
    \includegraphics[width=0.49\columnwidth]{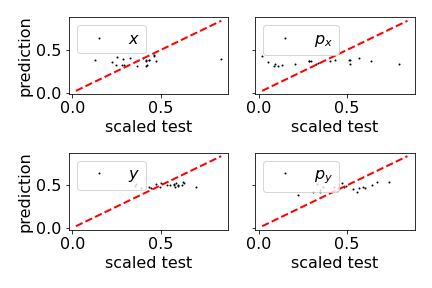}
    \includegraphics[width=0.49\columnwidth]{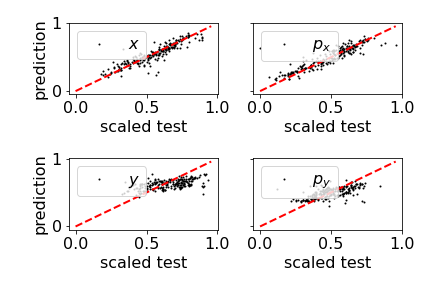}
    \includegraphics[width=0.49\columnwidth]{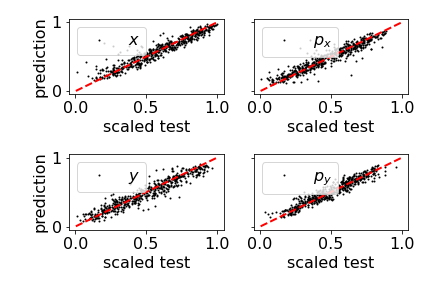}
    \caption{\label{fig:f_performance} Prediction accuracies of the ANN
      model in four different scenarios with different sextupole settings.
      Each subplot in every row (scenario) represents one of the projected
      dimensions ($x,p_x,y,p_y$), in which a black dot represents one
      particle's scaled coordinate. The dashed red line represents the
      desired value. Top left: a perfect surrogate model for a purely
      linear lattice (black dots and the dashed red lines overlap); Top
      right: a lattice with only chromatic sextupoles in place, the
      prediction accuracies are poor; Bottom left: improved accuracies are
      seen after introducing some harmonic sextupoles, but the settings
      are not optimal; Bottom right: a lattice with optimized harmonic
      sextupoles, the prediction accuracy is greatly improved.}
  \end{figure}

  Using the National Synchrotron Light Source II (NSLS-II) storage
  ring~\cite{nsls-ii2013} as an example, four special scenarios were
  studied to demonstrate the proof of concept, illustrated in
  Fig.~\ref{fig:f_performance}. First, consider a special scenario in
  which the lattice does not include any nonlinear magnets. The one-turn
  map is linear. In this case the model can accurately predict any testing
  trajectory.  As the system is a simple linear transformation, the level
  of accuracy is only limited by the computational numerical error, as
  shown in the first row of Fig.~\ref{fig:f_performance}.
  
  A linear lattice, however, cannot maintain stable beam at a high current
  due to its intrinsic negative chromaticity~\cite{lee2018}. The
  chromaticity represents a particle oscillation frequency shift due to
  its energy deviation. Therefore, some nonlinear sextupole magnets must
  be integrated into the lattice to compensate for the inherent
  chromaticity. In the second scenario of our example, the chromaticity
  was corrected to a value of positive 2 using chromatic sextupoles. After
  the correction, the same structured ANN model saw significantly reduced
  prediction accuracy. This was because particle motions became chaotic,
  and even unstable, when their trajectory diverged far from the
  origin. Due to strong nonlinearity, many particles fail to survive even
  for a single turn under these conditions. In the second scenario of
  Fig.~\ref{fig:f_performance}, low density clusters with large prediction
  discrepancies were observed. Only those survived test particles are
  plotted out.

  In the third scenario of Fig.~\ref{fig:f_performance}, some harmonic
  sextupoles were included to compensate for the nonlinearity caused by
  the chromatic sextupoles. These sextupoles were not fully optimized,
  therefore, limited improvements to the accuracy were seen. In the last
  scenario, after the harmonic sextupoles were optimized by minimizing the
  data-driven chaos indicator (as will be explained later), the chaos was
  significantly reduced and most simulated particles survived for at least
  a single turn. Consequently, the prediction accuracy of the ANN model
  was greatly improved in the last scenario, in which high density
  clusters were distributed around the desired prediction line.

  The correlations observed in the previous example indicate that the
  prediction accuracy of a surrogate model can be used as a chaos
  indicator for a complicated storage ring system. For a given surrogate
  model and a certain amount of training data, the model prediction
  becomes more inaccurate when particle motion in a lattice is more
  chaotic. Therefore, rather than constructing complex, high performance
  models, the prediction accuracy itself can be used as the objective for
  nonlinear lattice optimization. Another advantage is that the
  computational resources for the data analysis (for single turn
  trajectories) are greatly reduced, compared with an accurate DA
  computation through multi-turn (often thousands of turns) tracking.

\section{\label{sect:application}Applications in dynamic aperture optimizations}

  In this section the prediction accuracy of ANN models, used as a chaos
  indicator, was applied to optimize two different types of nonlinear
  lattices: the existing NSLS-II ring with double-bend achromat (DBA)
  lattice and a complex ESRF-EBS~\cite{farvacque2013} type hybrid
  multi-bend achromat (MBA) lattice.

  \subsection{\label{subsect:nsls2dba}NSLS-II's DBA Lattice}
\begin{figure}[!ht]
      \centering \includegraphics[width=0.5\columnwidth]{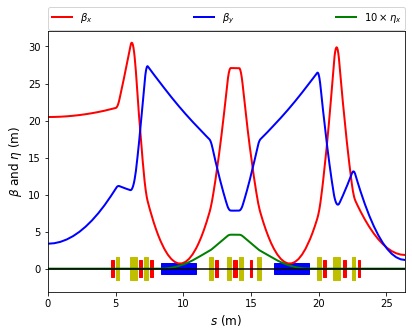}
      \caption{\label{fig:dbacell} Linear optics and magnet layout for one
        DBA cell of the NSLS-II storage ring. Red blocks represent
        sextupoles, blue ones are dipoles, and yellow ones are
        quadrupoles.}
    \end{figure}

    The NSLS-II~\cite{nsls-ii2013} is a dedicated $3^{\text{rd}}$
    generation medium energy (3 GeV) light source operated by Brookhaven
    National Laboratory. Its main storage ring lattice is a traditional
    DBA structure as illustrated in Fig.~\ref{fig:dbacell}. After the
    linear chromaticity is corrected by chromatic sextupoles, the
    available ``tuning knobs'' used for DA optimization are six families
    of harmonic sextupoles.

\begin{figure}[!ht]
      \centering
      \includegraphics[width=0.5\columnwidth]{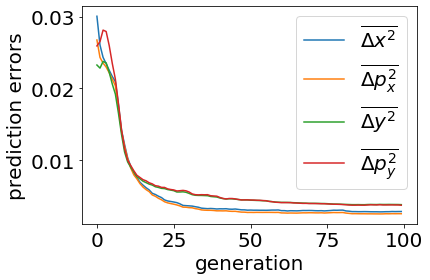}
      \caption{\label{fig:moga_convergency} Convergence of prediction
        accuracy measured with four mean squared errors of test datasets
        in the surrogate/MSE MOGA optimization.}
    \end{figure}

    A four-layer $4\times12\times12\times4$ ANN was chosen as the
    surrogate model to represent a one-turn transportation for on-momentum
    particles. The desired DA dimensions are $x=25\;mm$ and $y=10\;mm$ in
    the horizontal and vertical planes respectively, at the location of
    injection point. Thus, two elliptical areas in the phase space with
    axes at $(x,\frac{x}{\beta_x})$ and $(y,\frac{y}{\beta_y})$, are
    uniformly populated with 5,000 initial conditions as the input,
    $\overrightarrow{X}_0$. Here $\beta_{x,y}$ are the local Twiss
    parameters~\cite{courant1958}. The one-turn transportation can be
    accomplished with a symplectic particle tracking code, such as
    \textsc{elegant}~\cite{borland2000}. The coordinates at the exit are
    the output $\overrightarrow{X}_1$. The volume ratio of the training
    and testing data is 90\%:10\%. The python packages
    \textsc{scikit-learn}~\cite{scilearn2011} and
    \textsc{keras}~\cite{chollet2015} have been used for this
    application. To avoid over- or under-fitting, the maximum number of
    training epochs was set to a sufficiently large number, and an early
    stopping point was used to halt the training once the model
    performance ceased improving.  The mean squared error (MSE) in each
    phase space dimension is used as an independent optimization
    objective. By varying harmonic sextupole settings, the accuracy of the
    ANN model was monitored and used to drive a multi-objective genetic
    algorithm (MOGA) optimizer~\cite{deb2002}. Besides using the four MSEs
    in each dimension $(x,p_x,y,p_y)$ as the objectives, a minimum number
    of confined trajectories in the ensemble is used as a constraint. In
    other words, for a given sextupoles configuration, only when at least
    80\% of trajectories are confined within a predefined range
    $x,y\in[-0.1,0.1]\;m$, it can be considered as a qualified
    candidate. A good convergence of the average prediction accuracy was
    reached after a $100^{th}$ generation of evolution as shown in
    Fig.~\ref{fig:moga_convergency}. Using 100
    Intel\textsuperscript{\textregistered}
    Xeon\textsuperscript{\textregistered} 2.2-2.3 GHz CPU cores,
    optimization on this scale takes about 6 to 8 hours.

\begin{figure}[!ht]
      \centering
      \includegraphics[width=0.5\columnwidth]{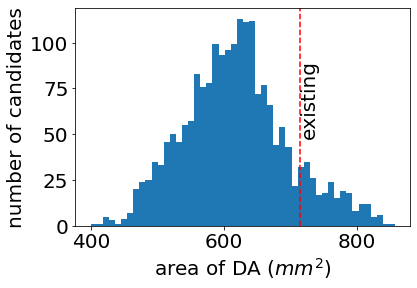}
      \caption{\label{fig:da_hist} Distribution of the DA areas in the
        last generation of the surrogate/MSE enhanced-MOGA
        optimization. The location of dashed line marks the DA area for
        the existing operation lattice~\cite{nsls-ii2013}}
    \end{figure}

    The DAs of the total population in the $100^{th}$ generation were
    calculated. The histogram of their DA areas in Fig.~\ref{fig:da_hist}
    shows that most of candidates have a sufficient DA, with the smallest
    ones being greater than 400 $mm^2$, which are more than sufficient to
    satisfy the requirements of machine operations. The existing operation
    lattice DA was also marked for a comparison.

\begin{figure}[!ht]
      \centering
      \includegraphics[width=0.5\columnwidth]{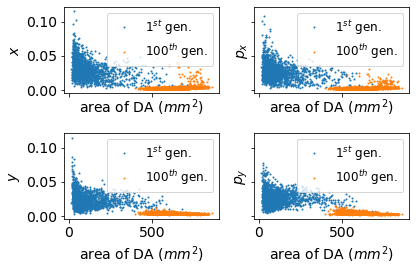}
      \caption{\label{fig:correlation} Correlation between the dynamic
        aperture and the ANN prediction accuracy seen in the MOGA
        optimization. The vertical axis in each subplot is the prediction
        accuracy in the corresponding dimension. The surrogate models of
        the $1^{st}$ generation (Blue dots) have low accurate predictions
        due to greater chaotic motion, and therefore, their original
        lattices have smaller DAs. Meanwhile, prediction accuracies of the
        $100^{th}$ generation are improved, as are their DAs.}
    \end{figure}

    To further confirm the correlation between the DA area and the ANN
    model's accuracy, The DA areas and average prediction accuracy for all
    candidates in the $1^{st}$ and $100^{th}$ MOGA generations were
    computed for comparison. As shown in Fig.~\ref{fig:correlation}, the
    settings for the $1^{st}$ generation have relatively poor model
    prediction accuracy, and their DAs are small. Meanwhile, prediction
    accuracy in the $100^{th}$ generation is significantly improved, as
    well as their DAs. Using this chaos indicator, which requires limited
    computation time and resources, significantly boosted the nonlinear
    lattice optimization.
  
\begin{figure}[!ht]
      \centering
      \includegraphics[width=0.5\columnwidth]{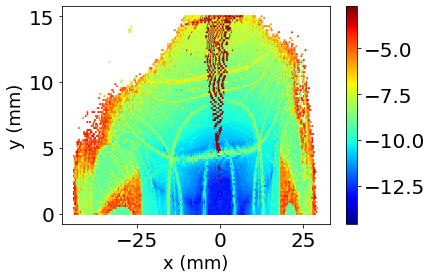}
      \caption{\label{fig:da_fma} On-momentum dynamic aperture colored
        with the diffusion, obtained from frequency map analysis for the
        candidate with the largest DA. The color represents the tune
        diffusion
        $\log_{10}(\Delta\nu_x^2+\Delta\nu_y^2)$~\cite{laskar1999}.}
    \end{figure}

\begin{figure}[!ht]
      \centering
      \includegraphics[width=0.5\columnwidth]{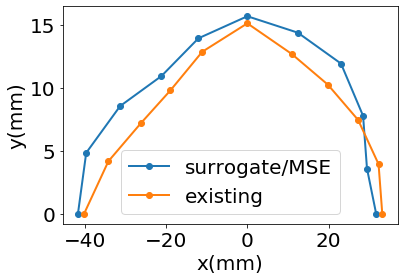}
      \caption{\label{fig:da_comparison} Comparison of on-momentum dynamic
        aperture between the best surrogate/MSE candidate (blue dots-line)
        and the existing operation lattice (yellow
        dots-line)~\cite{nsls-ii2013}.}
    \end{figure}

    The candidate with the largest on-momentum DA was chosen to implement
    a detailed frequency map analysis (FMA)~\cite{laskar1999} as shown in
    Fig.~\ref{fig:da_fma}. The DA comparison between this candidate and
    our existing operation lattice is shown in
    Fig.~\ref{fig:da_comparison}. An experimental test with live beam has
    been also carried out to confirm that this nonlinear lattice satisfies
    the requirements on the off-axis top-off injection and beam
    lifetime. Its DA is comparable to the solutions found using other
    methods~\cite{yang2011,li2016,li2018,li2021}.

\begin{figure}[!ht]
      \centering
      \includegraphics[width=0.5\columnwidth]{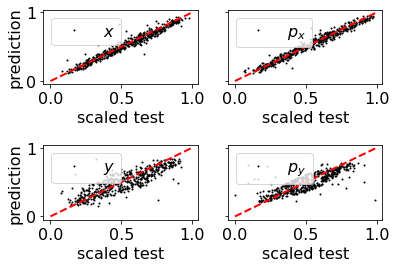}
      \includegraphics[width=0.5\columnwidth]{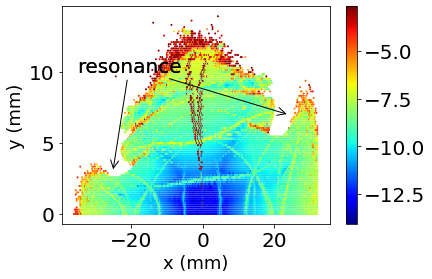}
      \includegraphics[width=0.5\columnwidth]{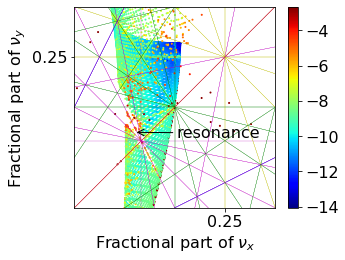}
      \caption{\label{fig:goodxbady} Top: good predictions in the
        horizontal plane, but poor ones in the vertical plane. Middle: DA
        crossing a resonance at small vertical amplitude can lead to
        particle loss due to more chaotic vertical motions. Bottom: Tune
        footprint in the tune space shows a large amplitude-dependent
        detuning in the vertical plane, which explains why the predictions
        in the vertical plane is worse.}
    \end{figure}

    Each chaos indicator in the 4-dimensional phase space is used as an
    independent and equally important optimization objective. Therefore,
    it is not surprising to find that some optimized results have
    different prediction accuracies in the horizontal and vertical
    planes. In the example illustrated in Fig.~\ref{fig:goodxbady}, the
    vertical predictions are significantly worse than the horizontal
    ones. A detailed DA tracking simulation shows that the vertical
    amplitude-dependent detuning is larger than the horizontal one, which
    leads to more chaotic vertical motions within the DA. When the tune
    footprint crosses $6^{th}$ order resonances even at a relatively small
    vertical amplitude $y\approx3\sim5\;mm$, particle loss can occur.
    Although this result does not provide a good DA solution, it shows
    that a data-driven chaos indicator is capable of distinguishing
    resonance-driven chaos among different dimensions of phase space, even
    for complicated dynamical systems such as a modern storage ring.

\begin{figure}[!ht]
      \centering
      \includegraphics[width=0.5\columnwidth]{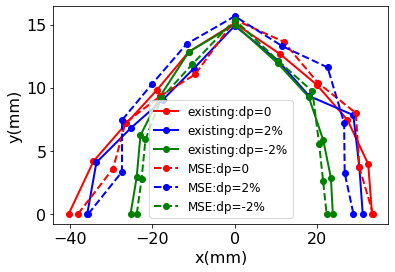}
      \caption{\label{fig:da5Dcomp} Comparison of on- and $\pm2\%$
        off-momentum DAs between the existing NSLS-II operation lattice
        (solid dot-lines) and the one optimized with surrogate/MSE in the
        5-dimensional phase space (dashed dot-lines). The sextupoles
        settings are listed in Tab.~\ref{tab:sext5DComp}.}
    \end{figure}

    Thus far, in the previous proof-of-principle study, only the
    on-momentum particles in the 4-dimensional transverse phase space are
    used for demonstration purposes. A sufficient off-momentum DA is of
    importance to ensure the beam lifetime. In an actual DA optimization,
    a 6-dimensional phase space should be considered. However, the
    longitudinal oscillation is relatively slow compared with the
    transverse ones. When designing a light source ring, usually a
    nonlinear lattice with fixed momentum-deviations are considered for DA
    optimization. In this case, the surrogate model's input have five
    components $\overrightarrow{X}_0=(x,p_x,y,p_y,\delta=\frac{\Delta
      p}{p_0})$, but still four components in the output because $\delta$
    is the same as its input. The settings for the MOGA optimization, such
    as the number of objectives and tuning knobs, are unchanged. Only the
    inputs of the ANN or polynominal's variables need to be expanded from
    four to five. The comparison of its on- and off-momentum DAs with the
    existing operation lattice is shown in Fig~\ref{fig:da5Dcomp}. The
    original and optimal sextupoles settings for the NSLS-II DBA lattice
    obtained with the 4- and 5-dimensional phase space are listed in
    Tab.~\ref{tab:sext5DComp}. Even the six families harmonic sextupoles
    ($SH$ and $SL$) settings are quite different, their DAs are still
    comparable. Since only single-turn tracking trajectories are used to
    train surrogate models to drive the MOGA optimizer, the computation
    cost of using the new technique is much lower.

    \begin{table}[ht]
      \caption{Sextupoles settings for existing lattice and obtained with
        4- and 5-D surrogate/MSE} \centering
      \begin{tabular}{|c|| r | r | r |} 
        \hline Name & existing $K_2\;(m^{-3})$ & $K_2\;(m^{-3})$ 4D &
        $K_2\;(m^{-3})$ 5D \\ [0.25ex] \hline\hline

        SH1 &   19.83291  &  20.49257 &  23.73833\\
        \hline
        SH3 &   -5.85511  &  -1.14901 &  -9.64248\\
        \hline
        SH4 &  -15.82090  & -21.40682 & -14.99301\\
        \hline
        SL1 &  -13.27161  &  -2.85597 & -17.44383\\
        \hline
        SL2 &   35.67792  &  26.11824 &  34.05922\\ 
        \hline
        SL3 &  -29.46086  & -24.72681 & -23.61317\\ 
        \hline
      \end{tabular}
      \label{tab:sext5DComp}
    \end{table}
    
  \subsection{\label{subsect:mba}Hybrid MBA Lattice}

\begin{figure}[!ht]
      \centering
      \includegraphics[width=0.5\columnwidth]{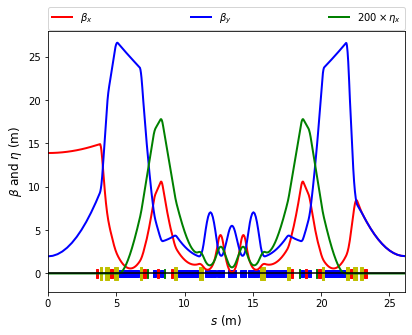}
      \caption{\label{fig:mbacell} Linear optics and magnet layout for one
        cell of an ESRF-EBS type~\cite{farvacque2013} MBA lattice. The
        lattice can provide a diffraction-limited horizontal emittance
        $30\;pm$ with a $3\;GeV$ beam. Red blocks are sextupoles, yellow
        ones are quadrupoles, and blue ones are either dipoles or off-axis
        quadrupoles (i.e., inverse bending magnets.)}
    \end{figure}

    Low-emittance light sources have entered a new era.  Various
    multi-bend achromat (MBA) type lattices have been developed to reach
    diffraction-limited horizontal emittances.  These lattices can deliver
    much brighter X-ray beams than seen before. In this section the
    data-driven chaos indicator was used to optimize the dynamic aperture
    of an MBA lattice type with seven bends, as shown in
    Fig.~\ref{fig:mbacell}.

\begin{figure}[!ht]
      \centering
      \includegraphics[width=0.5\columnwidth]{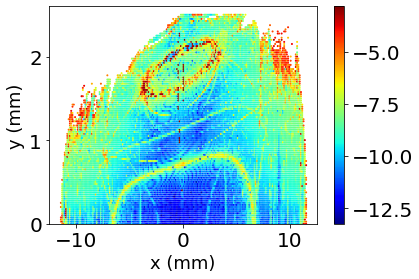}
      \caption{\label{fig:mbada} Dynamic aperture optimized with a
        data-driven chaos indicator for a complex MBA lattice. The color
        represents the tune diffusion.}
    \end{figure}

    In this example, several families of chromatic sextupoles were used.
    Maintaining constant linear chromaticities needs to be considered a
    constraint while tuning the sextupoles. Due to stronger sextupoles,
    the constructed surrogate model's prediction becomes worse than the
    previous DBA lattice. However, for an optimization problem, the chaos
    indicator only needs to provide relative values rather than absolute
    ones to guide the numerical optimizer to convergence. Similar
    convergence curves as Fig.~\ref{fig:moga_convergency} were observed
    but not shown here.  One of the optimal DAs is shown in
    Fig.~\ref{fig:mbada}, which is sufficient for adopting an off-axis
    injection scheme at one of its long straight sections. This example
    shows that the data-driven chaos indicator can still be applied to a
    complex lattice design.

\section{\label{sect:selection}Discussions on model selection and others}

  Two different surrogate models were used in our studies.  A
  $7^{th}$-order polynomial was used to study the H\'enon map.  In the
  examples provided by two storage ring lattices, a four-layer
  $4\times12\times12\times4$ ANN as shown in Fig.~\ref{fig:nn} was
  used. The selection of model complexity (the order in polynomial model,
  and the number of hidden layers and nodes in an ANN) compromise between
  model accuracy and processing time. In principle, a more complicated
  model can achieve predictions with greater accuracy, but would need more
  data and time to train. For these applications, high absolute prediction
  accuracy is not critical because only a relative accuracy is needed to
  direct the optimizer to converge. For example, a three-layer ANN was
  also tested and confirmed functional for the lattice
  optimization. However, its statistical fluctuations were larger than the
  one used in the four-layer ANN and therefore were not the focus of the
  investigation. Although, at the early stages of optimization, a course,
  simple-structured model could be used to quickly narrow down the ranges
  of searching parameters. This would be much more computationally
  efficient. Then another, more complex model can be deployed for a finer,
  more accurate and precise search. Besides the application of an ANN and
  a polynomial regression, another surrogate model using the support
  vector regression~\cite{smola2004} was also confirmed to be able to see
  the existence of correlations. Other surrogate models should be
  functional as well with this technique, although they were not tested in
  this investigation.

  There are also some other advanced machine-learning techniques available
  that can better evaluate accuracy of an ANN model. For example, using
  the K-fold cross validation method~\cite{bengio2004} ensures that every
  observation from the original dataset has the chance of appearing in
  training and test sets, especially when limited input data is
  available. In our examples for lattice optimization, rather than using
  the time-consuming cross-validation that K-fold requires, sufficient
  training data were generated with a particle tracking simulation code,
  because using only one-turn tracking for such rings requires much less
  of a demand on computational resources.

  The original H\'enon map and particle transformation are symplectic,
  i.e., they represent conservative Hamiltonian systems, while the
  surrogate models established from data are usually not. The lack of
  symplecticity can result in artificial damping or excitation in
  long-term transformations. The established surrogate models should not
  be used for the DA computation. However, while constructing data-driven
  chaos indicators, surrogate models were only used to characterize the
  sensitivity of transformations to their initial conditions. Even if the
  transformations are not perfectly symplectic, it does not affect such
  applications.

  Due to the lack of a real physics model, this chaos indicator cannot
  replace the diffusion rate obtained with the FMA, which measures the
  regularity of resonant motions of a nonlinear dynamical system. Using
  this indicator as the optimization objective might not be as competitive
  as the direct tracking-based optimization if not taking the computation
  cost into account. This method should be used as a complementary (rather
  than an alternative) tool for DA optimization. Because of its
  computation efficiency, a much larger population size can be used to get
  some potentially good solutions or narrow down the searching ranges
  quickly. Then some accurate but time-consuming tracking-based
  optimization can be further implemented.
  
\section{\label{sect:summary}Summary}

  A novel data-driven chaos indicator concept was introduced by
  correlating the degree of chaos of a dynamical system and its surrogate
  model's prediction accuracy. This indicator can be used to optimize the
  dynamic aperture of storage rings. Traditionally, the prediction
  accuracy of a model has been critically important for many
  machine-learning applications. With this method, however, the prediction
  accuracy is used as a relative indicator of the chaos of a dynamical
  system. Greater accuracy is therefore less important, and surrogate
  models which have a lower resource demand are sufficient for this
  purpose. This method also provides a new perspective on the
  characterization of chaos in nonlinear dynamical systems and an
  efficient method for dynamic aperture optimization.

\section*{Acknowledgements}
    
  We would like to thank Dr. Y. Hao (MSU) and the colleagues from NSLS-II
  for stimulating discussions. This research is supported by the
  U.S. Department of Energy under Contract No. DE-SC0012704 (BNL),
  National Natural Science Foundation of China (No. 11922512) and Youth
  Innovation Promotion Association of Chinese Academy of Sciences
  (No. Y201904). A. Liu is supported by Virginia Pond Scholarship Summer
  Intern Program at BNL (2019).

\bibliography{ddci.bib}

\end{document}